\documentclass[useAMS,usenatbib]{mn2e}

\usepackage{graphicx,amssymb}
\usepackage[normalem]{ulem}
\usepackage[usenames,dvipsnames]{xcolor}
\usepackage{soul}

\newcommand{\rev}{ }

\title[Forming only minor planets]
{Forming planetary systems that contain only minor planets}

\author[]{Dimitri Veras$^{1,2,3}$\thanks{E-mail: dimitri.veras@aya.yale.edu},
Shigeru Ida$^{4}$
\\
$^{1}$Centre for Exoplanets and Habitability, University of Warwick, Coventry CV4 7AL, UK
\\
$^{2}$Centre for Space Domain Awareness, University of Warwick, Coventry CV4 7AL, UK
\\
$^{3}$Department of Physics, University of Warwick, Coventry CV4 7AL, UK
\\
$^{4}$Earth-Life Science Institute, Tokyo Institute of Technology, Meguro, Tokyo 152-8550, Japan
}

\begin{document}
\label{firstpage}
\pagerange{\pageref{firstpage}--\pageref{lastpage}}
\maketitle

\begin{abstract}
Estimates of the frequency of planetary systems in the Milky Way are observationally limited by the low-mass planet regime. Nevertheless, substantial evidence for systems with undetectably low planetary masses now exist in the form of main-sequence stars which host debris discs, as well as metal-polluted white dwarfs. Further, low-mass sections of star formation regions impose upper bounds on protoplanetary disc masses, limiting the capacity for terrestrial or larger planets to form. Here, we use planetary population synthesis calculations to investigate the conditions that allow planetary systems to form only minor planets and smaller detritus. We simulate the accretional, collisional and migratory growth of $10^{17}$~kg embryonic seeds and then quantify which configurations with {\it entirely} sub-Earth-mass bodies ($\lesssim 10^{24}$~kg) survive. We find that substantial regions of the initial parameter space allow for sub-terrestrial configurations to form, with the success rate most closely tied to the initial dust mass. Total dust mass budgets of up to $10^2 M_{\oplus}$ within 10~au can be insufficiently high to form terrestrial or giant planets, resulting in systems with only minor planets. Consequently, the prevalence of planetary systems throughout the Milky Way might be higher than what is typically assumed, and minor planet-only systems may help inform the currently uncertain correspondence between planet-hosting white dwarfs and metal-polluted white dwarfs.
\end{abstract}


\section{Introduction}

The last 30 years of discoveries in exoplanetary science has revealed a vast diversity of planetary systems in mass, architecture, chemical composition and dynamical state.  Nevertheless, despite consistent technological advancements, the detection of the lowest mass planetary systems -- with sub-terrestrial bodies -- still usually resides beyond instrumental capabilities\footnote{Ironically, the first confirmed exoplanetary system contains sub-terrestrial planet masses \citep{wolfra1992,wolszczan1994}, and the first detected exoplanetary system does not yet feature any confirmed terrestrial or giant planets \citep{vanmaanen1917}.}.

The difficulty in directly detecting these low-mass systems has increasing consequences for pressing questions in the evolution of planetary systems. The first question is foundational: what is the true fraction of planetary systems in the Milky Way? Giant planet frequencies alone appear to be on the order of 10 per cent \citep{bownie2018,baretal2019,fuletal2021}. However,  when expanding the sample size to include super-Neptunes and mini-Neptunes, the frequency increases  to tens of per cent \citep{vacetal2024}. Then, when sampling even lower planetary masses in the form of terrestrial planets, the frequency may increase to beyond 50 per cent \citep{casetal2012,heetal2021}. What if we are able to probe even lower masses?

Another question relates to suggestive signatures in protoplanetary and debris discs orbiting main-sequence stars. All planets are formed in these discs. Nevertheless, despite the confirmations of over 5600 exoplanets which are not embedded in their birth discs, only a few exoplanets which actually are embedded in their birth discs have been detected with confidence \citep{kepetal2018,hafetal2019}. In dozens of other cases, planetary bodies represent a possible, but not definitive, explanation for a variety of disc signatures \citep{pinetal2023,pearce2024}. A more comprehensive assessment of planet formation would help assess the likelihood of these indirect signatures arising from planetary bodies.

A third question relates to white dwarf planetary systems. Over 1700 of these systems are known, almost entirely through the detection of planetary debris in the stars' photospheres \citep{wiletal2024}. How the debris arrives at the white dwarf remains an outstanding question, with an increasing number of potential dynamical avenues \citep[e.g.][]{trietal2022,ocoetal2023,akietal2024,pharei2024}, some of which do not feature any type of terrestrial or giant planetary perturber \citep{veretal2022}. Further, the known white dwarfs which are most likely to host exoplanets do not (yet) illustrate a clear correlation with detectable rocky debris in their stellar photospheres {\rev \citep{thoetal1993,sigetal2003,luhetal2011,ganetal2019,vanetal2020,blaetal2021,limetal2024,muletal2024,zhaetal2024}}.

All of these considerations motivate formation studies of planetary systems with only sub-terrestrial bodies. A potential starting point for such investigations is the true distributions of mass budgets in star forming regions and protoplanetary discs. However, as emphasised in the review article by \cite{draetal2023}, extracting the true dust mass from observations is difficult and uncertain \citep{muletal2021,mioetal2023}, and this extraction often leads to an underestimation of the true mass \citep{zhuetal2019}. As a result, performing a census of the mass budgets of the circumstellar environments of newly-born stars is fraught with challenges, and we do not attempt this task here.

Instead, we appeal to the simple argument that the masses of protoplanetary discs can be arbitrary low and that the lowest mass protoplanetary discs cannot be detected\footnote{For quantitative perspective, surveys of populations of protoplanetary discs in different star-forming regions estimate the amount of mass and pebbles to be $\sim 10^{-2}-10^3M_{\oplus}$ \citep{baretal2016,pasetal2016,lonetal2018,cieetal2019,tobetal2020,tycetal2020}. However, probing lower masses remains difficult at this time.}. Here, we focus on the 7-8 order-of-magnitude mass gap between the Earth ($\sim 10^{24}-10^{25}$~kg) and planetesimal seeds of mass\footnote{This $10^{17}$~kg value is less massive than the mass of each named moon in the solar system, and is about one order-of-magnitude less massive than the {\it Lucy} mission Trojan target 617 Patroclus.} $10^{17}$~kg. We assume that the seeds have already formed, and consider only their future evolution while embedded in a protoplanetary disc of gas and dust.

Our study is not the first to probe the sub-terrestrial mass regime from a theoretical point-of-view. \cite{najetal2022} evolve planetesimal rings with a total solid mass as low as $10^{-2} M_{\oplus}$, and show that minor planets were formed and survived disc dispersal. \cite{sanetal2024} investigate planet formation around low-mass stars, and show that core accretion efficiently forms minor planets. Our investigation features a different starting point -- with larger seeds -- as well as a different type of simulation code, and a dedicated parameter space exploration of this low-mass regime. In Section 2, we describe our methodology and the code used. Section 3 presents the results, and Section 4 contains a discussion. We summarise the work in Section 5, {\rev and show some additional plots for completeness in the Appendix}.

\section{Formation model}

In order to grow our $10^{17}$~kg seeds into minor planets, we use the core accretion model that has been developed over the last few decades by Ida \& Lin in a series of seven sequentially numbered papers (Papers I-VII) \citep{idalin2004a,idalin2004b,idalin2005,idalin2008a,idalin2008b,idalin2010,idaetal2013}. Many other core accretion prescriptions are of course available, including the more recent {\it New Generation Planetary Population Synthesis} models \citep{buretal2021,emsetal2021a,emsetal2021b,misetal2021,schetal2021a,schetal2021b}, plus models that specifically are intended to connect discs and planets \citep{buretal2022,emsetal2023}, and others which are not necessarily tied to population synthesis \citep[see e.g.][]{raymor2022}.

The oligarchic growth formulation from Paper VII \citep{idaetal2013} with a few tweaks suits our purposes because it relies on well-established formation physics; we are not attempting to reproduce modern or historic observations of exoplanetary systems or debris discs (as few such observations are available for the studied mass regime). Additionally, various details of giant planet formation, such as Type II migration and runaway gas accretion, are unimportant for our study.

This model (as explained in detail in Papers I-VII) assumes that the protoplanetary disc is parametrised by the following surface density dust distribution

\[
\Sigma_{\rm dust}\left(r\right)
=
\Sigma_{\rm d,10}\eta_{\rm ice}f_{\rm d}
       \left( \frac{r}{10 {\rm au}} \right)^{-q}
\]

\begin{equation}
\ \ \ \ \ \ \ \ \ \ \ \ \ \ 
=\Sigma_{\rm d,10}\eta_{\rm ice}f_{\rm g,10} 10^{\left[{\rm Fe/H}\right]}
       \left( \frac{r}{10 {\rm au}} \right)^{-q}
       \label{DustSigma}
\end{equation}

\noindent{}and the following surface density gas distribution

\[
\Sigma_{\rm gas}\left(r,t\right)
=
\Sigma_{\rm g,10}f_{\rm g}
       \left( \frac{r}{10 {\rm au}} \right)^{-p}
\]

\begin{equation}
\ \ \ \ \ \ \ \ \ \ \ \ \ 
=\Sigma_{\rm g,10}f_{\rm g,10} e^{-\frac{t}{\tau_{\rm dep}}}
       \left( \frac{r}{10 {\rm au}} \right)^{-p}.
       \label{GasSigma}
\end{equation}

\noindent{}In Eqs. (\ref{DustSigma})-(\ref{GasSigma}), $t$ refers to time and $r$ refers to distance from the star, which has a metallicity of [Fe/H]. The other parameters are all constants: $\Sigma_{\rm d,10}=0.32$ g cm$^{-2}$, $\Sigma_{\rm g,10}=75$ g cm$^{-2}$, $f_{\rm g,10}$ is a dimensionless scaling factor common to both distributions, $\eta_{\rm ice}$ accounts for the location of the snow line, $\tau_{\rm dep}$ is a proxy for the protoplanetary disc dissipation timescale through gas depletion, and $q$ and $p$ determine the steepness of dust and gas distributions, respectively.

We now elaborate more on these parameters. The values of $\Sigma_{\rm d,10}$ and $\Sigma_{\rm g,10}$ {\rev are normalisation factors that were used in Papers IV-VII and allow us here to model discs with surface densities within one order of magnitude of} the minimum mass solar nebula \citep[MMSN;][]{hayashi1981}. Hence, adjusting $f_{\rm g,10}$ by a few orders of magnitude below or above unity allows one to provide a rough comparison with the solar system's approximate nascent birth disc. The value of $\eta_{\rm ice}$ is taken to be 1.0 within the snow line, and 2.0 outside of the snow line\footnote{{\rev In Papers I-VII, the value of $\eta_{\rm ice}$ exterior to the snow line was taken to be 4.2 according to \cite{hayashi1981}, but it was recognised in Paper VII that this value may be smaller by a factor of 2 \citep{poletal1994}.}}. The location of the snow line itself, $a_{\rm snow}$, is taken to be $a_{\rm snow}=2.7~{\rm au}~\times \left(M_{\star}/M_{\odot}\right)$, where $M_{\star}$ is the mass of the star\footnote{{\rev This relation assumes a shallower mass dependence on luminosity than in Papers I-VII and is more reflective of pre-main-sequence stars \citep{stapal2004}. Regardless, this choice does not significantly affect the results, and for post-main-sequence systems, the consequences of this choice are negligible compared to the consequences of the star's luminosity increasing by many orders of magnitude during the giant branch phases.}}. For the other parameters, our fiducial values are $\tau_{\rm dep} = 10^{6.5}$~yr, $q=1.5$, and $p=1.0$, but here we explore the effect of the variation of these values on our results. 

Other parameters which are of particular interest are the available mass budgets for formation: the total initial mass of the disc, $M_{\rm disc}$, and the total initial dust mass, $M_{\rm dust}$. The latter is important because sub-terrestrial bodies will be rocky, and will retain little, if none, of the gas mass. 

Computation of these values requires us first to define the extent of our disc, from $a_{\rm min}$ to $a_{\rm max}$. Here we set $a_{\rm min} = 0.05$~au and $a_{\rm max} = 100$~au. Then we can compute:

\[
M_{\rm disc} \equiv M_{\rm disc, initial} \approx M_{\rm gas, initial}
\]

\[
\ \ \ \ \ \
= \int_{a_{\rm min}}^{a_{\rm max}} 2\pi r \Sigma_{\rm gas}\left(r,t=0\right)dr
\]

\[
\ \ \ \ \ \ 
=\frac{2 \pi \Sigma_{\rm g,10}f_{\rm g,10}}{\left(10 \ {\rm au} \right)^{-p}}
\left[
\frac{a_{\rm max}^{2-p} - a_{\rm min}^{2-p}}{2-p}\right], 
\ \ p \ne 2,
\]

\begin{equation}
\ \ \ \ \ \  
= \frac{2 \pi \Sigma_{\rm g,10}f_{\rm g,10}}{\left(10 \ {\rm au} \right)^{-2}}
\ln{\left( \frac{a_{\rm max}}{a_{\rm min}} \right)}, 
\ \ \ \ \ \ \ \ p = 2.
\label{Mdisc}
\end{equation}

\noindent{}In a similar manner, for the total dust mass, we have:

\[
M_{\rm dust} = \int_{a_{\rm min}}^{a_{\rm max}} 2\pi r \Sigma_{\rm dust}\left(r\right)dr
\]

\[
\ \ \ \ \ \ 
=\frac{2 \pi \Sigma_{\rm d,10}f_{\rm g,10}}{\left(10 \ {\rm au} \right)^{-q}}
1.0\times 10^{\left[{\rm Fe/H}\right]}
\left[
\frac{a_{\rm snow}^{2-q} - a_{\rm min}^{2-q}}{2-q}\right], 
\]

\[
\ \ \ \ \ \ \ \
+\frac{2 \pi \Sigma_{\rm d,10}f_{\rm g,10}}{\left(10 \ {\rm au} \right)^{-q}}
2.0\times 10^{\left[{\rm Fe/H}\right]}
\left[
\frac{a_{\rm max}^{2-q} - a_{\rm snow}^{2-q}}{2-q}\right], 
\ \ q \ne 2,
\]

\[
\ \ \ \ \ \  
= \frac{2 \pi \Sigma_{\rm d,10}f_{\rm g,10}}{\left(10 \ {\rm au} \right)^{-2}}
1.0 \times 10^{\left[{\rm Fe/H}\right]}
\ln{\left( \frac{a_{\rm snow}}{a_{\rm min}} \right)}
\]

\[
\ \ \ \ \ \ \ \
+ \frac{2 \pi \Sigma_{\rm d,10}f_{\rm g,10}}{\left(10 \ {\rm au} \right)^{-2}}
2.0 \times 10^{\left[{\rm Fe/H}\right]}
\ln{\left( \frac{a_{\rm max}}{a_{\rm snow}} \right)}, 
\ \ \ \ \ \ \ \  q = 2.
\]

\begin{equation}
\label{Mdust}
\end{equation}

We assume that $10^{17}$~kg seeds have already formed and are interspersed throughout the disc between $a_{\rm min}$ and $a_{\rm max}$. {\rev The seeds have the same mass as the background, non-individually resolved planetesimals, which are treated as a continuum. The planetesimal surface density is time-evolved due to embryo accretion, and is resolved through $10^3$ logarithmically-spaced radial bins (see Section 3.1 of Paper IV). Hence, the initial state represents a fluid of planetesimals with seed particles of the same mass. We set 3~g~cm$^{-3}$ for the rocky planetesimals and 1~g~cm$^{-3}$ for the icy planetesimals.  

The seeds interact with one another in two different ways depending on whether the gas disc is present, and our code self-consistently handles both. When the gas disc is present, the interaction is dictated by resonant interactions induced by orbital migration, and is modelled with an analytical prescription. This prescription is described in Section 2.4 of Paper VI. After the gas has been depleted, then the seeds, which have since grown into embryos, interact with each other through mutual impacts. These impacts are prompted by orbital eccentricity growth. The prescription for both eccentricity growth and mutual impacts is given in Section 2.6 and the Appendix of Paper VI, and is semi-analytic in nature. The application of this procedure has previously been compared favourably to outcomes of $N$-body simulations (Section 2.7 of Paper VI).
}

While the number of seeds varies per simulation depending on the initial conditions, this number is most strictly constrained by computational feasibility. In theory, one could maximally pack in $10^5$ or $10^6$ seeds between $a_{\rm min} = 0.05$~au and $a_{\rm max} = 100$~au. However, doing so is computationally unrealistic. Here, we place a strict upper limit of, at most, $10^3$ seeds per simulation, and importantly ensure that they are sufficiently separated so that the full width of their feeding zones do not overlap. {\rev We discuss the implications of this necessary restriction in Section 4.3.}

\begin{figure*}
\centerline{\Huge \underline{A fiducial case}}
\centerline{}
\centerline{}
\centerline{
\includegraphics[width=17cm]{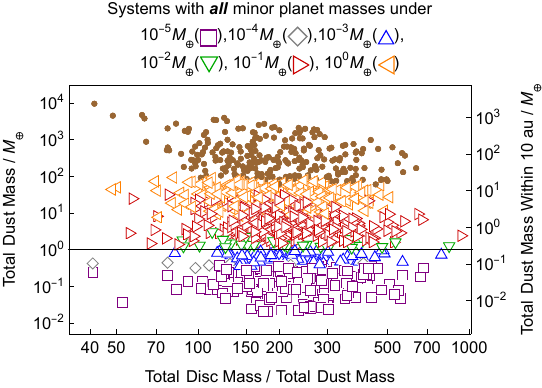}
}
\centerline{}
\centerline{
\includegraphics[width=8.5cm]{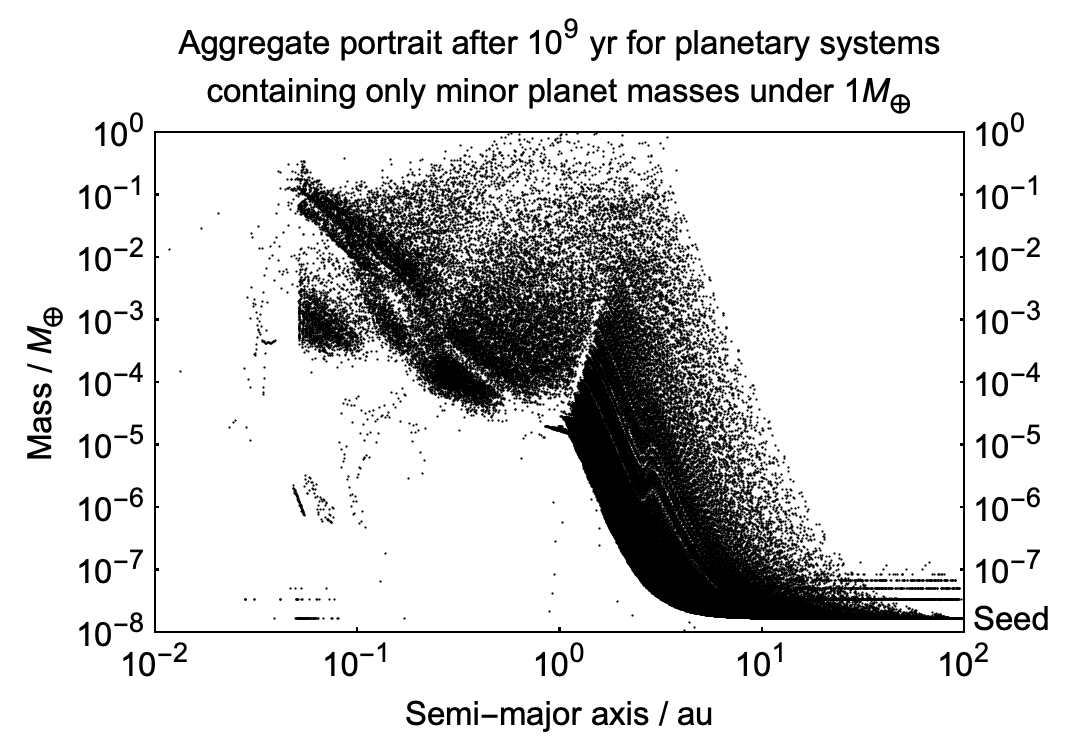}
\includegraphics[width=8.5cm]{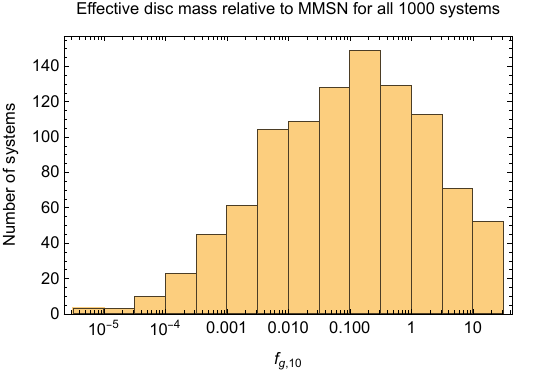}
}
\caption{
Forming planetary systems with only sub-Earth-mass minor planets, denoted {\it sub-terrestrial systems}. In the top panel, each symbol represents a different planetary system; the coloured shape symbols are the sub-terrestrial systems, and the filled brown circles indicate systems within which at least one surviving body formed with a mass $\ge 1M_{\oplus}$. The lower left panel contains the final state of only the sub-terrestrial systems, with the mass of the seeds indicated on the right $y$-axis, and the lower right panel illustrates the distribution of $f_{\rm g,10}$ that was used to generate all systems, where MMSN denotes Minimum Mass Solar Nebula. In all cases, $q=1.5$, $p=1.0$, $M_{\star} = 1.0M_{\odot}$ and $\tau_{\rm dep} = 10^{6.5}$~yr.
}
\label{FigFiducial}
\end{figure*}

\begin{figure*}
\centerline{\Huge \underline{Effects of gas and dust distributions for $M_{\star}=0.5M_{\odot}$}}
\centerline{}
\centerline{}
\centerline{
\includegraphics[width=9.5cm]{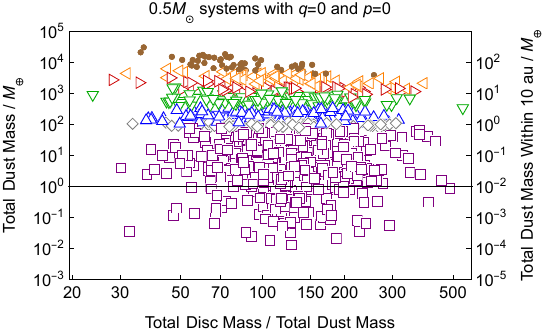}
\includegraphics[width=9.5cm]{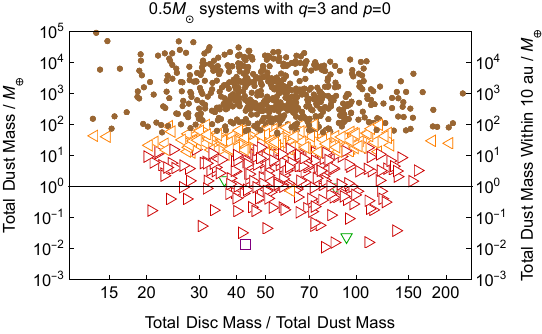}
}
\centerline{}
\centerline{
\includegraphics[width=9.5cm]{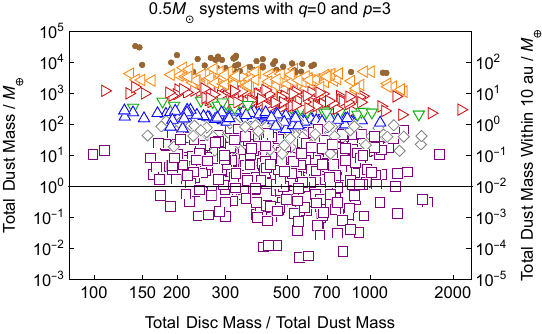}
\includegraphics[width=9.5cm]{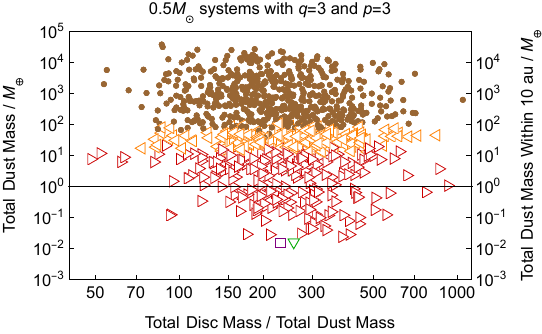}
}
\caption{
Dependence of sub-terrestrial system formation on the initial distributions of dust and gas, through $q$ and $p$, for host stars of $M_{\star}=0.5M_{\odot}$. The meaning of the symbols are the same as in Fig. \ref{FigFiducial}. The values of $q$ and $p$ shown were chosen to sample extremes, and they bound a realistic range of initial dust and gas distributions. This figure demonstrates that sub-terrestrial system formation is a strong function of $q$, and occurs in a widespread fashion as long as the dust mass is sufficiently low. 
}
\label{FigM05}
\end{figure*}

\begin{figure*}
\centerline{\Huge \underline{Effects of gas and dust distributions for $M_{\star}=2.0M_{\odot}$}}
\centerline{}
\centerline{}
\centerline{
\includegraphics[width=9.5cm]{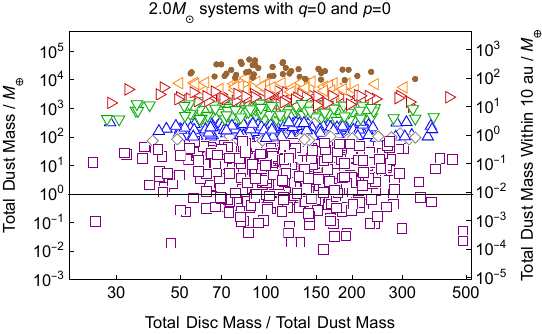}
\includegraphics[width=9.5cm]{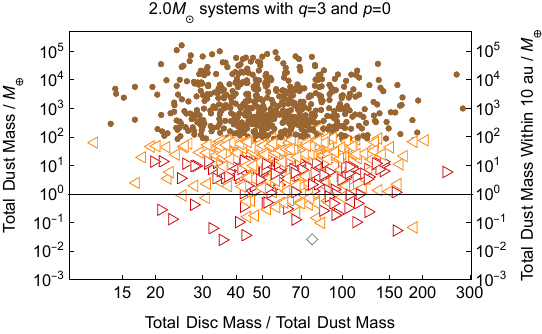}
}
\centerline{}
\centerline{
\includegraphics[width=9.5cm]{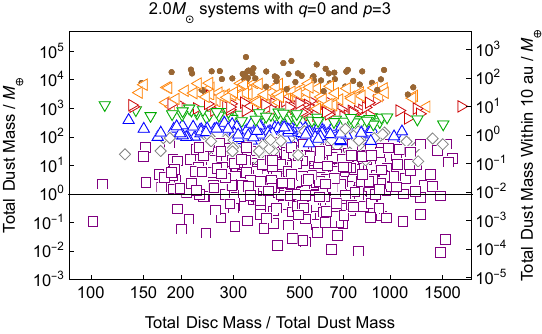}
\includegraphics[width=9.5cm]{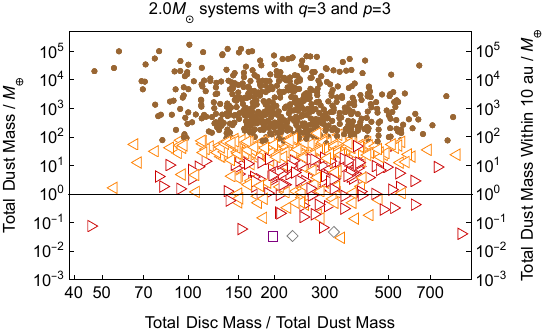}
}
\caption{
Same as Fig. \ref{FigM05}, but for $M_{\star}=2.0M_{\odot}$. Comparison of this figure with Fig. \ref{FigM05} showcases few qualitative differences, the largest being the vertical dispersion in the different sets of triangles in the right panels. Overall, sub-terrestrial system formation is not directly affected much by stellar mass.
}
\label{FigM20}
\end{figure*}

\begin{figure*}
\centerline{\Huge \underline{Effects of disc dispersal timescales}}
\centerline{}
\centerline{}
\centerline{
\includegraphics[width=9.5cm]{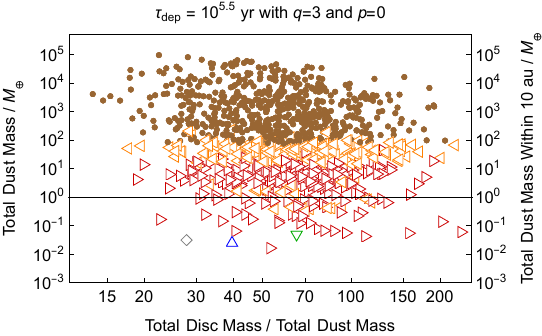}
\includegraphics[width=9.5cm]{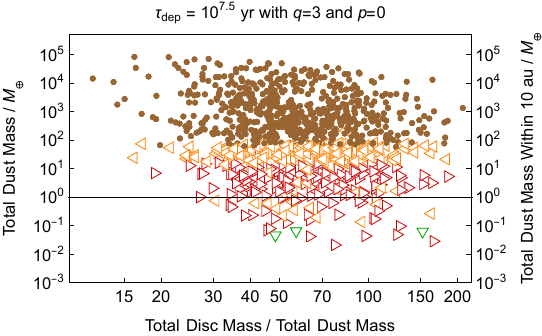}
}
\centerline{}
\centerline{
\includegraphics[width=9.5cm]{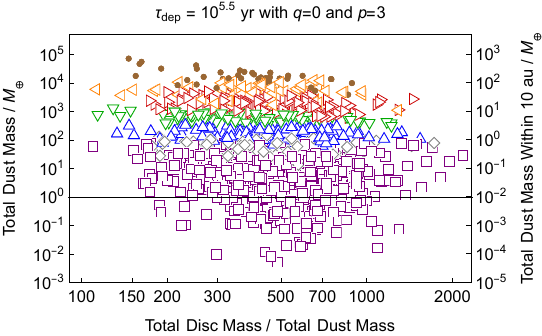}
\includegraphics[width=9.5cm]{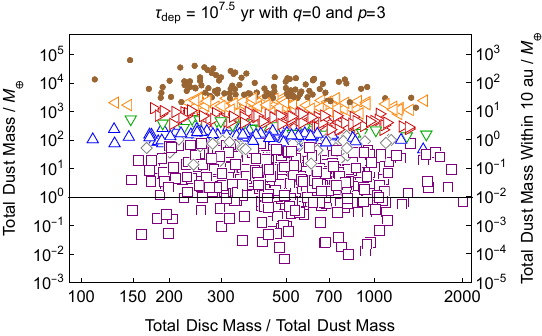}
}
\caption{
How the protoplanetary disc dispersal timescale, $\tau_{\rm dep}$, affects the formation of sub-terrestrial systems, for host star masses of $1M_{\odot}$. The meaning of the symbols are the same as in Fig. \ref{FigFiducial}. The left panels feature $\tau_{\rm dep} = 10^{5.5}~{\rm yr}\approx 0.32~{\rm Myr}$ and the right panels feature $\tau_{\rm dep} = 10^{7.5}~{\rm yr}\approx 32~{\rm Myr}$, whereas in Figs. \ref{FigFiducial}-\ref{FigM20}, $\tau_{\rm dep} = 10^{6.5}~{\rm yr}\approx 3.2~{\rm Myr}$. The top plots sample steep dust profiles ($q=3$) and show almost no dependence on $\tau_{\rm dep}$, and the bottom plots sample flat dust profiles ($q=0$), whose formation outcomes are more strongly dependent on $\tau_{\rm dep}$.
}
\label{Figtdep}
\end{figure*}

\begin{figure*}
\centerline{\Huge \underline{Sub-terrestrial-only formation for given dust budgets}}
\centerline{}
\centerline{}
\centerline{
\includegraphics[width=7.5cm]{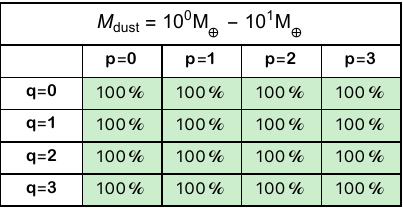} \ \ \ \ \ \ \
\includegraphics[width=7.5cm]{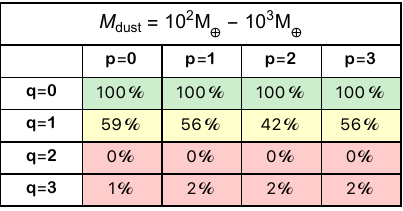}
}
\centerline{}
\centerline{
\includegraphics[width=7.5cm]{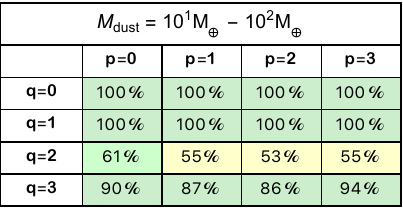} \ \ \ \ \ \ \
\includegraphics[width=7.5cm]{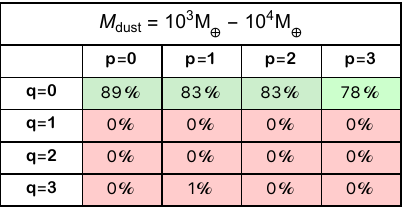}
}
\caption{
Efficiency with which dust mass budgets can produce sub-terrestrial systems. Each percentage represents the fraction of systems sampled that formed only minor planets all with masses under $1M_{\oplus}$. Each table represents a different order-of-magnitude range of initial dust mass and includes results from 16 sets of simulations according to the given values of $q$ and $p$. To calculate how much of the dust mass is within 10~au only, multiply the total dust mass by $\left\lbrace 0.0096, 0.087, 0.59, 0.99 \right\rbrace$ for $q=\left\lbrace 0,1,2,3\right\rbrace$. The colour scheme corresponds to 0-20 per cent (red), 20-40 per cent (pink), 40-60 per cent (yellow), 60-80 per cent (lime) and 80-100 per cent (green). The figure demonstrates that total dust mass budgets of $\sim 10^{0}-10^{2}M_{\oplus}$ commonly produce sub-terrestrial systems, although dust budgets of $10^{4}M_{\oplus}$ can do so too when $q=0$.
}
\label{FigTables}
\end{figure*}

\begin{figure*}
\centerline{\Huge \underline{Consequences for post-main-sequence evolution}}
\centerline{}
\centerline{}
\centerline{
\includegraphics[width=9.5cm]{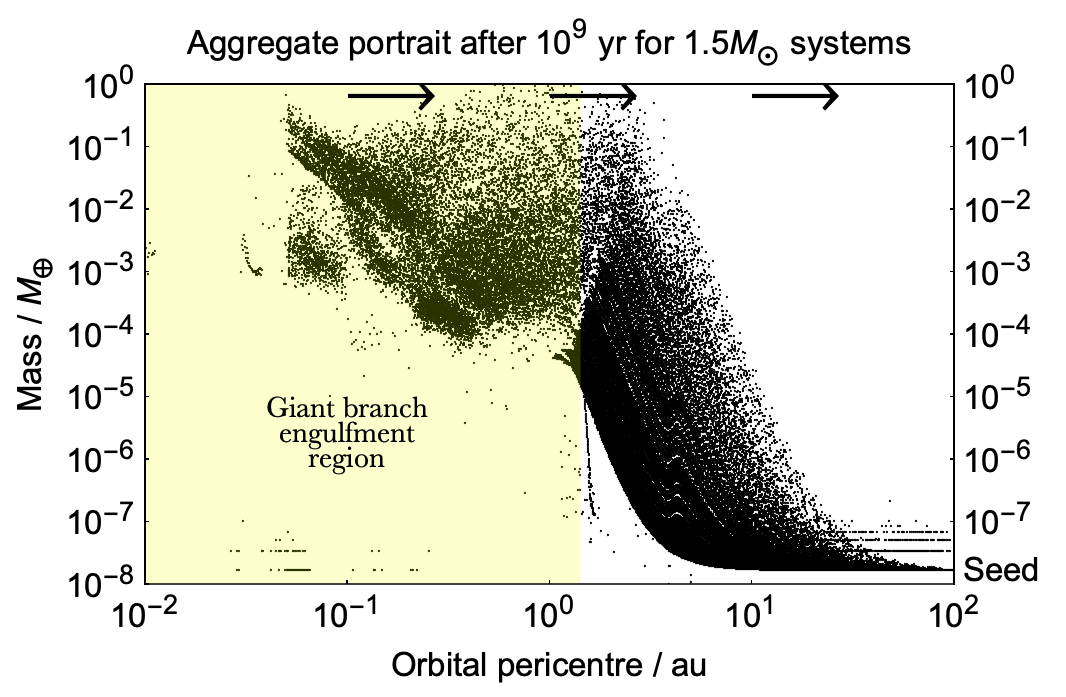}
\includegraphics[width=9.5cm]{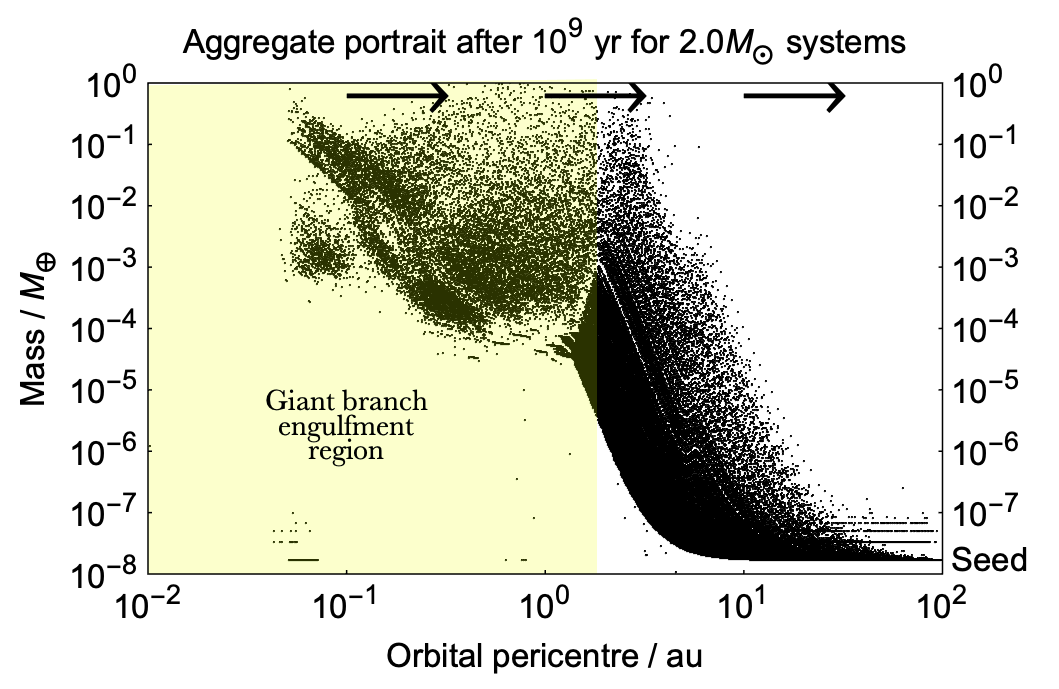}
}
\centerline{}
\centerline{
\includegraphics[width=9.5cm]{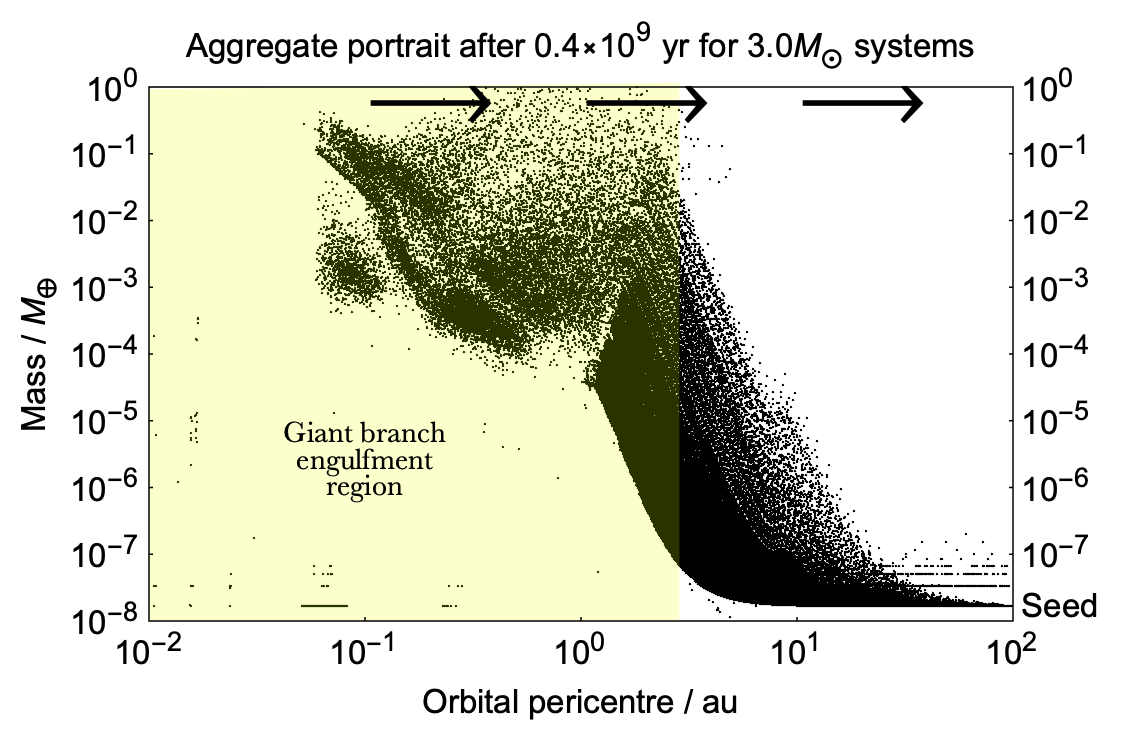}
\includegraphics[width=9.5cm]{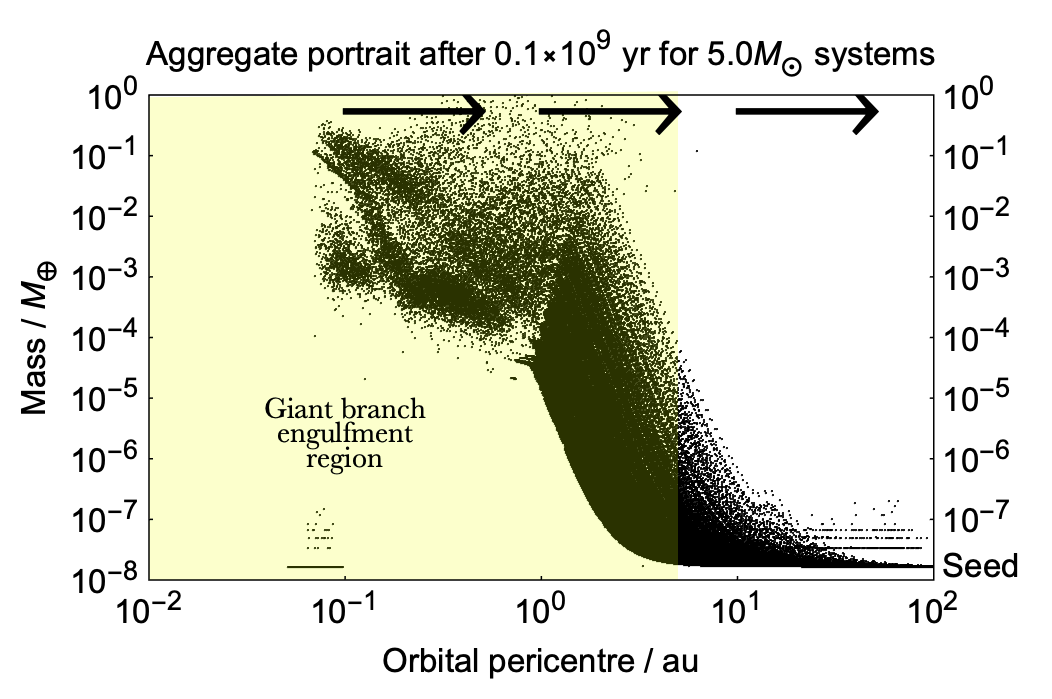}
}
\caption{
Resulting minor planet system architectures for different stellar masses, and approximate locations where sub-terrestrial systems would be subjected to stellar engulfment from their parent star's giant branch evolution. In all systems, $q=1.5$, $p=1.0$ and $\tau_{\rm dep}=10^{6.5}$~yr, and the stopping time of the simulation was selected to ensure that it did not exceed the star's main-sequence lifetime. The arrows indicate, to scale, the extent of the orbital expansion of the minor planets due to stellar mass loss during giant branch evolution. The plots illustrate that the higher the stellar mass, the less likely that surviving minor planets in sub-terrestrial systems will avoid engulfment, and those which do survive will harbour the lowest minor planet masses; many of these will be unprocessed seeds.
}
\label{FigPostMS}
\end{figure*}

\section{Results}

Having established the parameters for the model, we now proceed to construct sets of simulations which allow us to explore sub-terrestrial system formation. Of particular interest is determining if and under what conditions {\it only} sub-terrestrial bodies can form {\it and} survive after a protoplanetary disc has dissipated. Consequently, we present our results in a way which highlights these systems.

In order to probe the most relevant parameter space, we perform many sets of simulations, where each set consists of 1,000 individual simulations. Within an individual set, we fix $M_{\star}$, $q$, $p$ and $\tau_{\rm dep}$, but vary $f_{\rm g,10}$ and [Fe/H]. This choice allows us, within every set, to both sample a variety of magnitudes of both $M_{\rm disc}$ and $M_{\rm dust}$, as well as ensuring that their ratio is not fixed amongst different simulations. For each individual simulation, $f_{\rm g,10}$ is selected from a random distribution in log space with a mean of -1.0 and a dispersion of 2.0 (allowing us to focus on low-mass systems, but still sampling high mass ones and very low mass ones) and [Fe/H] is selected from a random distribution in linear space with a mean of 0.0 and a dispersion of 0.3 (allowing for the entire range of known exoplanet host star metallicities to be sampled).

Our choices for the fixed values per simulation set depends on the intended application. For $M_{\star}$, we consider $0.5M_{\odot}$, $1.0M_{\odot}$, $1.5M_{\odot}$, $2.0M_{\odot}$, $3.0M_{\odot}$ and $5.0M_{\odot}$. The lowest of these values is very common in the Milky Way; the Sun-like value adheres to the human desire to seek out solar system analogues; and the higher values are particularly relevant for the known white dwarf planetary systems, as will be discussed later. Because the main-sequence lifetime of each of these stars differs, we end our simulations at $10^9$~yr for $M_{\star} = 0.5M_{\odot}, 1.0M_{\odot}$, $1.5M_{\odot}$ and $2.0M_{\odot}$, but at sooner times for more massive stellar hosts. For $\tau_{\rm dep}$, our default value is $10^{6.5}$~yr~$\approx 3.2$~Myr, but we will sample the effect of altering this value by one order of magnitude in both directions.

Regarding the distribution of dust and gas within the disc, {\rev we adopt $q=1.5$ for the dust and $p=1.0$ for the gas (corresponding to} a steady accretion disc with a constant $\alpha$ viscosity). However, we allow for both much flatter and steeper gradients. Hence, we sample values in the ranges $q=0-3$ and $p=0-3$. These ranges yield $M_{\rm disc}/M_{\rm dust} \sim 10^0 - 10^4$, which is sensible given that the canonical gas-to-dust ratio in protoplanetary discs is $10^2$.

\subsection{Fiducial case}

First, we illustrate the outcome of a fiducial set of simulations, in Fig. \ref{FigFiducial}. In this set, we assume $q=1.5$, $p=1.0$, $M_{\star} = 1.0M_{\odot}$ and $\tau_{\rm dep} = 10^{6.5}$~yr. The top plot of the figure contains one data point per system and showcases two types of outcomes after 1 Gyr of simulation time. 

The first type are simulations where at least one body with a mass at least as large as $1M_{\oplus}$ formed and survived protoplanetary disc dispersal. These systems are shown as filled brown circles, and are not of particular interest here, except to provide a useful visual boundary for the other systems.

The second type are the systems of greatest interest, denoted by large coloured shape symbols. In these systems, sub-terrestrial bodies formed and survived and have masses which are all below the value indicated by the symbol. These are the systems to which we henceforth refer as {\it sub-terrestrial systems}. Amongst these, the systems with the smallest bodies are illustrated by the purple squares, and the systems with the largest bodies are illustrated by the leftward-pointing orange triangles. Any symbol appearing below the thin black horizontal line must represent a sub-terrestrial system.

In this particular, fiducial, case, the upper plot demonstrates that for typical gas/dust (or disc/dust) ratios, sub-terrestrial systems are easily formed when the total {\rev initial} dust mass is up to two  orders of magnitude greater than an Earth mass. In fact, the correlation of the formation of these systems with dust mass is strong, as indicated by how the different symbols appear in roughly horizontal bands. The right $y$-axis displays the dust mass budget within 10~au only, where the seeds predominately reach their isolation masses.

The lower left panel of Fig. \ref{FigFiducial} shows the aggregate spatial and mass distributions for all sub-terrestrial systems superimposed on the same plot. The spatial range is selected as such to demonstrate (i) the initial seed mass (right y-axis) relative to minor planet masses up to $1M_{\oplus}$, (ii) the outer extent of the disc ($10^2$~au), and (iii) the inner semi-major axis at which particles are assumed to be lost through stellar-induced destruction ($10^{-2}$~au). 

This plot illustrates several aspects of the simulations. First, the horizontal lines between $10^{-7}M_{\oplus}$ and $10^{-8}M_{\oplus}$ indicate largely unaccreted and unprocessed seeds. Second, the highest mass minor planets, close to $1M_{\oplus}$, tend to form around $a_{\rm snow}$ (at about 2.7~au here). Third, outwards of $\sim 10$~au, minor planet masses remain low (under $\sim10^{-4}M_{\oplus}$). Fourth, inwards of $10^{-1}$~au, few minor planets migrate and/or survive in this region. In the context of Papers I-VII, all bodies in this plot would be considered ``rocky" except for the outermost seeds, which would instead be considered ``icy"; none of these objects would be considered ``gaseous".

The lower-right plot of Fig. \ref{FigFiducial} indicates how we sampled simulations for this figure with our distribution of $f_{\rm g, 10}$ (which represents a proxy multiplicative factor for $\left[1.4\times~{\rm MMSN}\right]$). This distribution includes all 1,000 simulations sampled; a priori, we did not know which would eventually form sub-terrestrial systems.

\subsection{Dependence on distribution of dust and gas}

We next consider the effect of altering $q$ and $p$ on sub-terrestrial minor planet formation. In order to bound these effects, we adopt extreme limits of both $q$ and $p$ for $M_{\star} = 0.5M_{\odot}$ stars in Fig. \ref{FigM05} and for $M_{\star} = 2.0M_{\odot}$ stars in Fig. \ref{FigM20}. For all simulation sets, we adopt $\tau_{\rm dep} = 10^{6.5}$~yr.

Both figures clearly demonstrate that (i) minor planet formation is a stronger function of $q$ than $p$, and that (ii) sub-terrestrial systems can form for any combination of physically reasonable values of $q$ and $p$. As $q$ is decreased while keeping all other variables fixed, the maximum dust mass which enables the formation of sub-terrestrial systems increases, and up to a value of $\sim10^4M_{\oplus}$ for the entire disc. However, in these cases, most of the mass is in the outer part of the disc. When comparing the right and left $y$-axes of both Figs. \ref{FigM05} and \ref{FigM20}, one can see that independent of $q$, the maximum dust mass budget within 10~au that can produce sub-terrestrial systems is about $10^{2}M_{\oplus}$.

In each of Figs. \ref{FigM05} and \ref{FigM20}, the $y$-axis ranges are fixed. However, the $x$-axis ranges differ in all 8 plots. Hence, sub-terrestrial system formation is not strongly sensitive to the disc's gas-to-dust ratio, but instead is strongly sensitive to $q$.

The differences between Figs. \ref{FigM05} and \ref{FigM20} appear to be overall slight. The most significant difference is the dispersion in maximum minor planet mass as a function of dust mass (compare the interleaving of the rightwards-pointing red triangles with the leftwards-pointing yellow triangles in the $q=3$ plots). Also, for the $p=3$ cases, the $0.5M_{\odot}$ systems allow for the formation of at least one planet more massive than $1M_{\oplus}$ with lower dust masses (by just $\sim 10$ per cent) than in the $2.0M_{\odot}$ systems. These differences showcase the weak direct dependence of sub-terrestrial system formation on host star mass.

\subsection{Dependence on disc dispersal timescale}

Figs. \ref{FigFiducial}-\ref{FigM20} all assume that $\tau_{\rm dep} = 10^{6.5}$~yr $\approx 3.2$~Myr. Now, we test how sensitive our results are on this choice, especially since disc dispersal timescales vary system-to-system. 

In Fig. \ref{Figtdep}, we sample simulation outcomes with both $\tau_{\rm dep} = 10^{5.5}$~yr $\approx 0.32$~Myr and $\tau_{\rm dep} = 10^{7.5}$~yr $\approx 32$~Myr. These values span what is assumed to be the disc dispersal timescales for the vast majority of stellar systems \citep{mamajek2009,ercetal2011,ribetal2015,lixia2016,monetal2023}. In order to sample extremes, we have also chosen $q=3$ and $p=0$ in the top panel plots and $q=0$ and $p=3$ in the bottom panel plots. We adopt $M_{\star}=1M_{\odot}$ for all these simulations.

The simulation outputs shown in the figure reinforces intuition that a longer-lived disc allows for higher-mass minor planets to form. In other words, for a given dust mass, the upper limit on minor planet mass is higher for $\tau_{\rm dep} = 10^{7.5}$~yr than for $\tau_{\rm dep} = 10^{5.5}$~yr, as can most easily be seen on the plots around dust mass bands of $10^2-10^4M_{\oplus}$ for $q=0$. However, the overall effect of this two orders-of-magnitude shift of disc dispersal timescale on sub-terrestrial system formation is slight, and instead is likely more important for terrestrial (and giant) planets.

\subsection{Formation frequencies as a function of dust mass}

So far, we have shown that the maximum mass at which minor planets are formed in sub-terrestrial systems is strongly correlated with the initial dust mass budget. In fact, in most of the plots in Figs. \ref{FigFiducial}-\ref{Figtdep}, the same shape symbols appear in near-horizontal bands on the plots.

Consequently, we now provide more detailed statistics of formation frequency as a function of dust mass band, as well as $q$ and $p$ in increments of 1.0. We perform 16 sets of 1,000 simulations with $M_{\star}=1M_{\odot}$ and $\tau_{\rm dep}=10^{6.5}$~yr, and report our results in Fig. \ref{FigTables}.

The figure consists of 4 frequency tables, one for each order-of-magnitude range in dust mass. Each box contains the percentage of all initial condition sets sampled at the given values of (i) dust mass band, (ii) $p$ and (iii) $q$ for which sub-terrestrial systems were formed. The colour coding corresponds to: 0-20 per cent (red), 20-40 per cent (pink), 40-60 per cent (yellow), 60-80 per cent (lime) and 80-100 per cent (green).

The figure suggests that the most efficient total dust mass budget for forming sub-terrestrial systems is $10^0-10^2M_{\oplus}$. Lower dust mass budgets always generate these systems, and flatter dust density profiles can produce these systems for the highest dust mass budgets. The figure reports total dust mass budgets; in order to calculate how much of the dust mass is within 10~au only, multiply the total dust mass by $\left\lbrace 0.0096, 0.087, 0.59, 0.99 \right\rbrace$ for $q=\left\lbrace 0,1,2,3\right\rbrace$. The dip in ubiquitous low-mass formation frequency for $q=2$ and $M_{\rm dust} = 10^0-10^2M_{\oplus}$ can be attributed to this dust distribution clustering around $a_{\rm snow}$, allowing seeds to more easily accrete $>1M_{\oplus}$ of mass (whereas the dust clusters inwards of $a_{\rm snow}$ for $q=3$, and does not cluster much anywhere for $q=1$ or $q=0$).

\subsection{Beyond the main sequence}

The potential survival of sub-terrestrial systems beyond the main sequence has potential consequences for white dwarf planetary systems, which predominately feature rocky planetary debris rather than planets themselves. Although simulating these systems during the giant branch phases of evolution is beyond the scope of this work given the complexities involved in modelling the radiative effects \citep{veretal2019,feretal2022,lietal2024}, we can nevertheless at least estimate which minor planets would be engulfed by the expanding host star.

To do so, we perform an additional four sets of 1,000 simulations, where each set has $q=1.5$, $p=1.0$, and $\tau_{\rm dep}=10^{6.5}$~yr, but with different stellar masses and stopping times $t_{\rm stop}$. The host star mass is particularly important because the most common progenitor mass for the current white dwarf planet population is about $2M_{\odot}$ \citep{cumetal2018,cunetal2024}. Further, the highest progenitor mass for a white dwarf planetary system is about $5M_{\odot}$ \citep{holetal2021}.

Hence, for our four sets of simulations, we adopt $M_{\star} = 1.5, 2.0, 3.0, 5.0M_{\odot}$. The higher the stellar mass, the shorter the main-sequence lifetime, $t_{\rm ms}$, and in each case this lifetime should not significantly exceed $t_{\rm stop}$. In order to compute both $t_{\rm ms}$ and the maximum stellar radius $R_{\rm max}$ during post-main-sequence evolution, we use the {\tt SSE} stellar evolution code \citep{huretal2000}. For $M_{\star} = \left\lbrace 1.5, 2.0, 3.0, 5.0 \right\rbrace M_{\odot}$, we obtain $t_{\rm ms} = \left\lbrace 2.6, 1.2, 0.4, 0.1 \right\rbrace$~Gyr, and $R_{\rm max} = \left\lbrace 1.4, 1.8, 2.9, 5.0 \right\rbrace$~au. Hence, we adopt $t_{\rm stop} = \left\lbrace 1.0, 1.0, 0.4, 0.1 \right\rbrace$~Gyr.

We illustrate the results of the simulations in Fig. \ref{FigPostMS} in plots of masses versus orbital pericentre for sub-terrestrial systems. The orbital pericentre is more important than the semi-major axis with regards to the future survival of the bodies in order for them to avoid engulfment within the expanding star. However, the vast majority of the minor planets in our systems are formed on and remain on nearly circular orbits. Hence, plots of mass versus semi-major axis would look very similar.

Each plot is partitioned into two regions. The left region, which has a background colour of yellow, approximates the maximum extent of the star during the giant branch phase through the value of $R_{\rm max}$. This number roughly corresponds to the maximum value which is acquired on the asymptotic giant branch phases of stellar evolution, which exceeds the maximum value on the red giant branch phases for these types of stars \citep{musvil2012,adablo2013}. 

Minor planets in this region would actually not necessarily be engulfed because their orbital expansion due to stellar mass loss might allow these objects to ``outrun" the expanding stellar envelope \citep{hadjidemetriou1963}. We show with three arrows near the top of each plot the extent of this orbital expansion under the adiabatic assumption \citep{veretal2011}: for $M_{\star} = \left\lbrace 1.5, 2.0, 3.0, 5.0 \right\rbrace M_{\odot}$, the orbital expansion factors are $\left\lbrace 2.6, 3.1, 4.0, 5.0 \right\rbrace$.

The right region in each plot, without a background colour, contains minor planets which are not in danger of engulfment, but instead will be subjected to radiative sublimation, melting, YORP and Yarkovsky effects \citep{jurxu2010,juretal2012,veretal2014,veretal2015,veretal2019,versch2020,malper2016,malper2017a,malper2017b,katz2018,lietal2024}. Hence, by no means will these minor planets be ``safe", but they will not be directly engulfed by the star. Further, they will not be ejected from the system, given the lack of larger perturbers. Hence, these minor planets will remain in some, perhaps broken-down, form, until the white dwarf phase of stellar evolution.

The potential consequences for white dwarf planetary systems will be discussed in {\rev Section 4.5}. Here, we just comment on what the figure illustrates. First, the symbol distribution on all four plots look very similar, demonstrating the weak dependence of sub-terrestrial system formation both on stellar mass and the ending time of the simulations. Second, based simply on $R_{\rm max}$, higher mass stellar hosts will be left with lower-mass minor planets. In the $5.0M_{\odot}$ case, these minor planets all will have masses under $10^{-4}M_{\oplus}$.

\section{Discussion}

We now discuss various aspects and implications of our results.

\subsection{Minor planet migration}

Our simulations do incorporate both Type I and II migration. However, the latter is irrelevant for our studied mass regime, and the former has a small-to-negligible effect on the results. Paper VII shows that the Type I migration timescale scales as $\sim 10^5 (M_{\rm seed}/M_{\oplus})^{-1}~{\rm yr}$, where $M_{\rm seed}$ is the mass of the seed. Hence, just for this timescale to be comparable to the disc depletion timescale, the seed would need to accrete enough mass to reach $\sim 10^{-1}M_{\oplus}$. Even just that accretion process takes time ($\sim 3\times 10^5 f_{\rm d}^{-1}f_{\rm g}^{-2/5}~{\rm yr}$), during which the disc is consistently depleting.

\subsection{Isolation mass}

Given that migration is typically ineffectual, a seed should remain at a relatively fixed orbital distance from the star accreting material in its feeding zone, with width $\Delta a_{\rm c}$, until either all the material in this zone has been depleted or until the disc dissipates. In the former case, the final mass of the seed becomes the so-called {\rev ``local isolation mass", $M_{\rm iso}$, given by

\[
M_{\rm iso} \approx 2 \pi r \Sigma_d \Delta a_{\rm c}
\]
\[
\ \ \ 
\approx 20 \pi a^2 \left( \frac{2M_{\rm iso}}{3M_{\star}} \right)^{1/3}
\Sigma_{\rm d,10}\eta_{\rm ice}f_{\rm g,10} 10^{\left[{\rm Fe/H}\right]}
       \left( \frac{r}{10 {\rm \ au}} \right)^{-q}
\]


\begin{equation}       
\ \ \ 
\sim 0.9 M_{\oplus} \left(\frac{a}{10~{\rm au}} \right)^{3-\frac{3q}{2}}
f_{\rm g,10}^{\frac{3}{2}}\left[ \eta_{\rm ice} 10^{\left[{\rm Fe/H}\right]} \right]^{\frac{3}{2}}      
,
\label{Miso}
\end{equation}
}

\noindent{}where the term in square brackets typically does not vary by more than one order of magnitude.

This local isolation mass represents the upper mass limit for fully-formed minor planets. For our simulations, the term in parentheses can change the numerical coefficient of the equation by a {\rev factor of $\sim 10^{-7}-10^{2}$}, and the value of $f_{g,10}$ can change {\rev that factor by $\sim 10^{-8}-10^3$}.

Although in this paper we have presented ensembles of simulations, equation (\ref{Miso}) can be useful to place a bound on an individual system. It also shows an explicit dependence of the upper mass limit on $q$ and $a$, helping to explain the trends seen in the simulation output.

However, equation (\ref{Miso}) should be used just as a guide. In our simulations, we do not artificially truncate planetesimal accretion according to this equation. The evolution of the surface density distribution of planetesimals by accretion of seeds is self-consistently calculated in the code (see Section 2.2 of Paper VI). As a result, in-situ growth of a seed naturally stops at the isolation mass: the local planetesimal surface density is depleted as the seed mass approaches the isolation mass.  

In not all cases does accretion allow a seed to reach the local isolation mass. In fact, in the outer regions of the disc, planetesimal growth is slow. As a result, many partially accreted or largely unaccreted seeds will remain behind after the disc dissipates. Examples of these largely unprocessed seeds can be seen in Figs. \ref{FigFiducial} and \ref{FigPostMS}, and they may represent a source of white dwarf planetary signatures, as described {\rev in Section 4.5}.

{\rev
\subsection{Computational limitations}

The mass of the seeds, coupled with the requirement that their feeding zones do not overlap, has both theoretical and computational consequences. The smaller the seed mass, the more of the formation process that could be modelled, and the greater number of seeds that could be packed into a given radial range. However, computationally, the practical upper limit of the number of seeds which could be modelled by our code is approximately $10^3$.

Consequently, there should be a larger reservoir of icy and rocky material in the outer regions of planetary systems than predicted in this paper. If no planets form in these regions to scatter this material away, then the small bodies will be dynamically stagnant and contribute to the already substantial low-mass tail of objects seen in the aggregate portraits of Figs. \ref{FigFiducial} and \ref{FigPostMS}. The consequences for post-main-sequence systems may be significant, as this more substantial reservoir could allow for higher accretion rates of planetary metals onto white dwarfs at a wider variety of cooling ages in the absence of planets themselves \citep{veretal2022}.

Potential improvements to our model, aside from the application of increased computational power, include incorporating analytical integration of the initial planet formation stage in order to facilitate seed placement \citep{voeetal2020,voeetal2021a,voeetal2021b} or from more general, purely analytic mass scaling arguments and estimates \citep[e.g.][]{goletal2004,emsetal2023}.

}

\subsection{Post-formation gravitational instability}

The results of Fig. \ref{FigPostMS}, and their comparison to Fig. \ref{FigFiducial}, indicate that subsequent to disc dissipation, sub-terrestrial planetary architectures remain relatively static. Without any terrestrial or giant planets to dynamically excite the minor planets, the minor planets can only excite one another.

However, they may be too far separated from one another to do so. $\Delta a_{\rm c}$ is approximately equal to 10 mutual Hill radii. The required number of mutual Hill radii to induce instability over long timescales is unclear because both it is a function of the minor planet-to-star mass ratio and because of a lack of dedicated investigations at very low ratios. This ratio is typically explored for terrestrial and giant planet masses but not often for minor planets \citep{chaetal1996,zhoetal2007,chaetal2008,smilis2009,funetal2010,puwu2015,yeeetal2021,ricste2023,yanetal2023}.

In addition to gravitational scattering, another potential avenue for instability is escape from the system. This prospect can be quantified with the Safronov number \citep{safronov1972}

\[
\Theta = \frac{a}{R_{\rm p}}\left(\frac{M_{\rm p}}{M_{\star}} \right)
\]

\[
\ \ \ \
\approx 5 \times 10^{-4} \left(\frac{M_{\rm p}}{10^{-3}M_{\oplus}} \right)^{2/3}
                         \left(\frac{M_{\star}}{M_{\odot}} \right)^{-1}
                         \left(\frac{a}{1~{\rm au}}\right)
                         \left(\frac{\rho_{\rm p}}{2~{\rm g/cm}^3} \right)^{1/3} 
\]

\begin{equation}
\ \ \ \ 
\ll 1
\end{equation}

\noindent{}where $M_{\rm p}$, $R_{\rm p}$ and $\rho_{\rm p}$ represent the mass, radius and density of the minor planet. Escape becomes likely when $\Theta \gtrsim 1$. However, as can be seen here, even for high minor planet masses, $\Theta$ remains less than one. Hence, escape from the system is unlikely.

\subsection{Post-main-sequence evolution}

As a result of minor planets not escaping the system and probably not scattering off of one another, their orbits will remain fixed for the main-sequence lifetime of their parent star. Then, the minor planets which avoid engulfment during the giant branch phases (Fig. \ref{FigPostMS}) will stick around even longer, until the star becomes a white dwarf.

However, these minor planets will remain in the system in a possibly broken-up or partially sublimated form, depending on a variety of radiative effects \citep{jurxu2010,juretal2012,veretal2014,veretal2015,veretal2019,versch2020,malper2016,malper2017a,malper2017b,katz2018,lietal2024}. Any dust which is greater than the blow-out size will survive to the white dwarf phase \citep{bonwya2010,donetal2010,zotver2020}, albeit potentially severely displaced due to radiative migration \citep{veretal2015,veretal2019}. The dust, boulders, asteroids and minor planets which then orbit the white dwarf may reach the stellar photosphere through radiation alone \citep{veretal2022} or through mutual gravitational scatterings, but only for the highest-mass minor planets \citep{verros2023}.

Neither of these pathways requires terrestrial or giant planets to shepherd material to the white dwarf photosphere\footnote{Planetary debris that the white dwarf accretes is often {\rev referred} to as ``metal pollution" because it produces observational signatures in otherwise pristine atmospheres composed of only H and He \citep{bonsor2024,xuetal2024}.}. However, the viability of either mechanism to reproduce accretion rates of pollution onto white dwarfs strongly depends on both the amount of mass which survives -- which our simulations cannot quantify given that they are resolution-limited with respect to the initial number of seeds -- and the amount of time in which this material resides in debris discs surrounding the white dwarfs \citep{giretal2012,verhen2020}. 

\begin{figure}
\includegraphics[width=8cm]{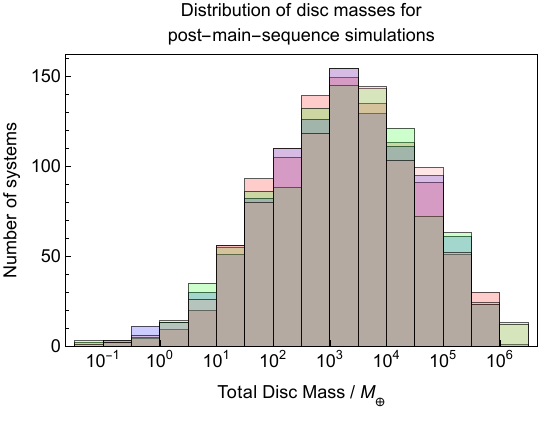}
\caption{
{\rev
The total disc masses of all four ensembles of systems that were simulated to generate each panel of Fig. \ref{FigPostMS}. The brown bars indicate where all four ensembles overlap, and the other colours are different ensemble overlap permutations. This plot illustrates that all four underlying disc distributions that were simulated are similar. An accurate accounting of the amount of mass retained in these systems throughout the white dwarf phase would require the incorporation of the true correlation of disc properties with stellar mass (coupled with radiative modelling during the giant branch phases).
}
}
\label{MassHist}
\end{figure}

{\rev
Further, obtaining a realistic assessment of the amount of material which survives would entail accounting for how disc properties change with the main-sequence stellar host mass. Observations suggest that host star mass and disc mass are positively correlated \citep{pasetal2016,ansetal2017,tobetal2020}, as is host star mass and accretion rate, acting as a proxy for gas disc mass \citep{manetal2023}. As for $\tau_{\rm dep}$, the value of this parameter is thought to decrease with higher stellar masses \citep{ribetal2015,ricetal2018}. 

None of these trends are reflected in Fig. \ref{FigPostMS}, in which all of the simulations use the same underlying distribution of disc properties. To demonstrate that, we overplot the ensemble of disc masses adopted in each of those simulations on top of one another in Fig. \ref{MassHist}. The differences in the four different distributions are relatively minor compared to the expected trends with stellar mass. As more metal-polluted white dwarfs with progenitor masses exceeding a few solar masses are discovered \citep{ouldetal2024}, dedicated planet formation studies in intermediate-mass and high-mass discs will become increasingly important \citep{veretal2020,kunetal2021,johetal2024}.
}

So far, none of the white dwarfs which show robust indications of orbiting planets have detected metal pollution {\rev 
\citep{thoetal1993,sigetal2003,luhetal2011,ganetal2019,vanetal2020,blaetal2021,limetal2024,muletal2024,zhaetal2024}.
} However, the reason may be due to observational limitations. Hence, claiming that there is an anti-correlation between planet hosts and metal polluted hosts would be too strong at this time. Nevertheless, as observations reveal more information, a comparison of  different pollution mechanisms, such as the ones presented in this section, become increasingly warranted.

{\rev
\subsection{Model extrapolation to observed systems}

The ability of our code to match theoretical predictions with observations -- the essence of population synthesis -- has been demonstrated in Papers I-VII. For this paper, enough observations of exclusively sub-terrestrial systems do not yet exist for a similar comparison to be conducted. 

Nevertheless, illustrating that the code remains valid in terms of broadly reproducing aspects of the exoplanet population helps improve the predictive power of our study. To this end, we have retained information about the subset of simulations which did form planets, and show this information in Figs. \ref{MassFig1} and \ref{MassivePostMS}.

These figures reveal that our code roughly matches key aspects of the modern (as of 2024) known exoplanet population \citep{cloutier2024}: (i) the abundant formation of Super-Earths and Mini-Neptunes, which are now thought to represent two of the most common classes of exoplanets in the Milky Way, (ii) the formation of gas giants, and (iii) the presence of all these classes of planets within $\sim 0.1$~au, indicating that inward migration (whether it be Type I, II or III migration) is common. 

One aspect of the exoplanet population which has revealed itself more fully since the publication of Paper VII but has actually been predicted by the code in some sense since Paper I is the so-called ``Neptunian desert" \citep[e.g.][]{fraetal2023,guietal2023,hawetal2023,osbetal2023,feretal2024,nabetal2024}. Fig. \ref{MassFig1} clearly reproduces a semi-major axis (or orbital period) deficit which is reflective of this desert. 

Another feature of the exoplanet population which has become apparent over the last decade is the ``radius gap" or ``radius valley" \citep[e.g.][]{fulpet2018,petetal2022}. This gap is not reproduced by the code, perhaps because the code does not incorporate prescriptions for atmospheric evaporation. However, atmospheric evaporation is not relevant for minor planets. Two other, more longstanding aspects of the exoplanet population that the code has not been able to reproduce (since Paper I) are the pile-up of gas giants beyond 1~au and their relatively high orbital eccentricities.

Finally, our code generates super-Earths and mini-Neptunes in the $1-10$~au range. These planets are currently difficult to detect due to observational biases. However, the {\it Nancy Grace Roman Space Telescope} is predicted to open up this key parameter space to observations. In this respect -- without digressing too much from the main message of our study -- our code predicts that these planets are formed, migrate and settle in this region.
}

\section{Summary}

We have investigated the formation of planetary systems that contain {\it only} minor planets with masses $\sim 10^{-7}M_{\oplus}-10^{0}M_{\oplus}$ by simulating the accretional, collisional and migratory growth of $10^{17}$~kg seeds embedded within a nascent protoplanetary disc. We find that:

\begin{itemize}

\item These systems are readily formed throughout the parameter space of initial conditions, suggesting that they may contribute substantially to estimates of planetary system frequency throughout the Galaxy.

\item The mass of the largest minor planets in these systems is most strongly correlated to the initial dust mass of the disc, and next most strongly correlated to the spatial distribution of this dust through the power-law exponent $q$.

\item These systems can form from initial dust mass budgets of up to $\sim 10^{2}M_{\oplus}$ within 10~au.

\item The minor planets in these systems which reside beyond about 10~au rarely accrete enough mass to exceed about $10^{-4}M_{\oplus}$. In this region, for flat distributions of dust, a substantial number of mostly unaccreted seeds will remain behind after the protoplanetary disc dissipates. {\rev The true number of these dynamically stagnant seeds may be even higher that what has been modelled here due to computational limitations.}

\item The formation of these systems are weakly dependent on the stellar mass {\rev (when assuming equal disc masses)}, disc dissipation timescale, gas density profile, Type I migration effects, and duration of the simulations.

\item The outermost minor planets in these systems may survive both main-sequence and giant branch evolution, and end up as a potential pollution reservoir for white dwarfs without any assistance from terrestrial or giant planets. 

\end{itemize}

\section*{Acknowledgements}

{\rev We thank the expert reviewer for thoughtful and constructive comments that have improved the manuscript}. This research was supported by the Munich Institute for Astro-, Particle and BioPhysics (MIAPbP), which is funded by the Deutsche Forschungsgemeinschaft (DFG, German Research Foundation) under Germany's Excellence Strategy – EXC-2094 – 390783311.

\section*{Data Availability}

The numerical implementation of the model presented in this paper is available upon reasonable request to the second author.

{\rev
\appendix
\renewcommand{\thefigure}{A-\arabic{figure}}
\section*{Appendix}

Although the focus of this work is on sub-terrestrial bodies, some of the simulations we ran did generate planets with masses greater than $1M_{\oplus}$. Partly for completeness and partly to demonstrate the workings of the population synthesis code, we show the planets that were formed from the simulations which were carried out to generate the plots in Figs. \ref{FigFiducial} and \ref{FigPostMS} in Figs. \ref{MassFig1} and \ref{MassivePostMS}. The figures demonstrate an expected outcome of sampling a wide range of $f_{\rm g,10}$ values: in many cases super-Earths are formed, and in some cases, giant planets are too. Just as with the minor planets, most of the larger planets will be engulfed by the host star during the giant branch phases.

\begin{figure}
\includegraphics[width=8cm]{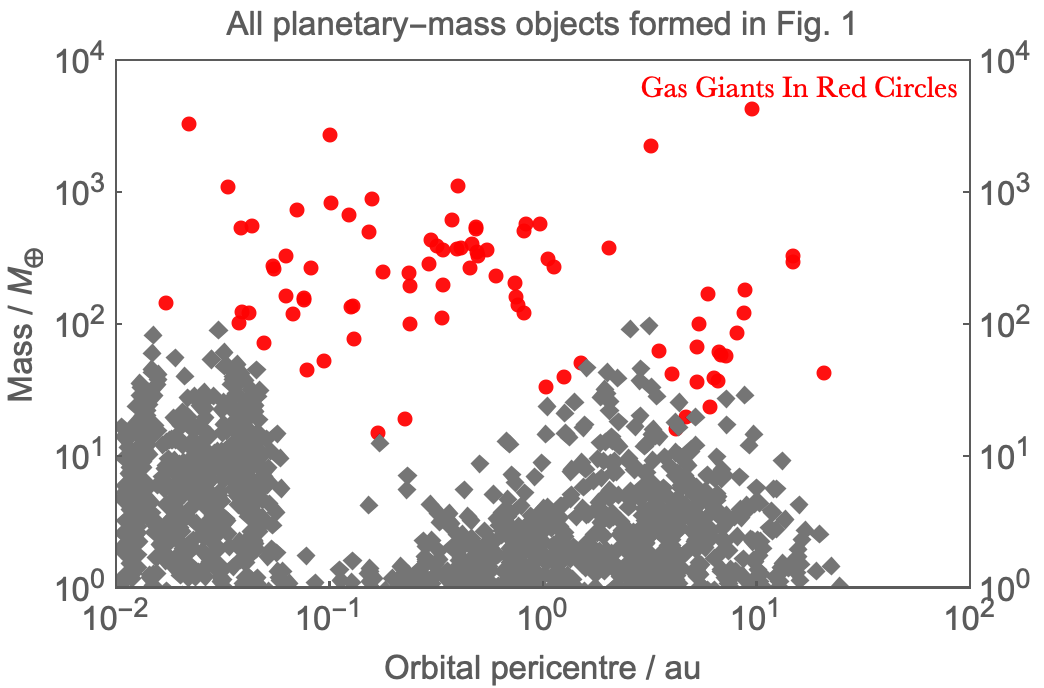}
\caption{
{\rev
The planetary-mass objects formed in the simulations whose output is displayed in Fig. \ref{FigFiducial} (corresponding to the filled brown circles in the top panel of that figure). In this figure, the red circles indicate gas giants and the black diamonds indicate other planets.
}
}
\label{MassFig1}
\end{figure}

\begin{figure*}
\centerline{\Huge \underline{Planetary-mass survivors of post-main-sequence evolution}}
\centerline{}
\centerline{}
\centerline{
\includegraphics[width=9.5cm]{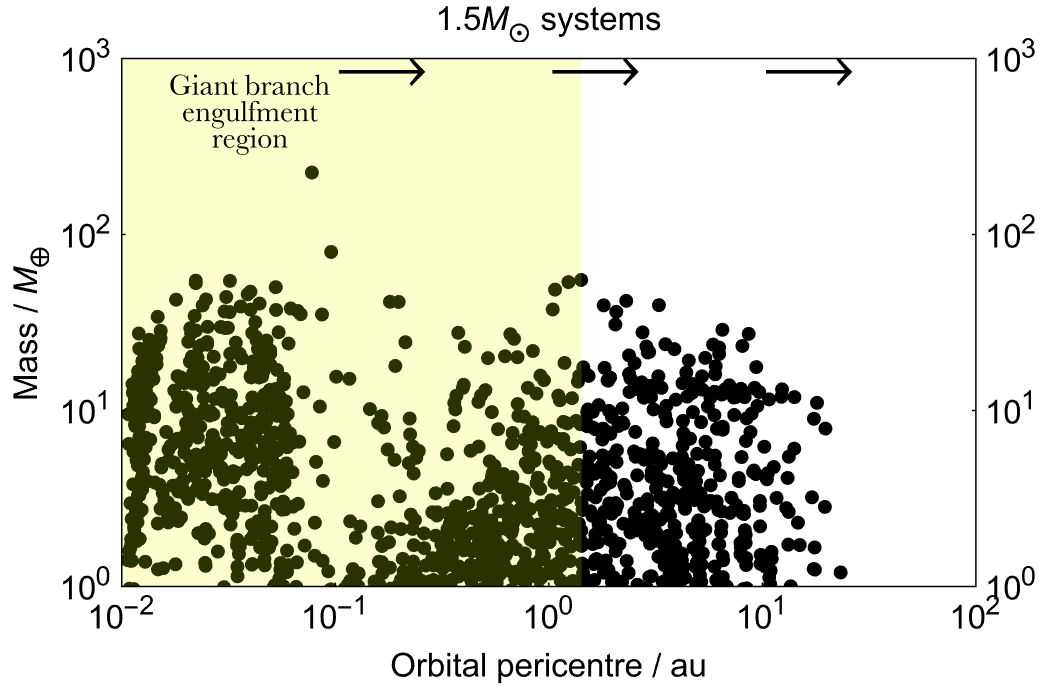}
\includegraphics[width=9.5cm]{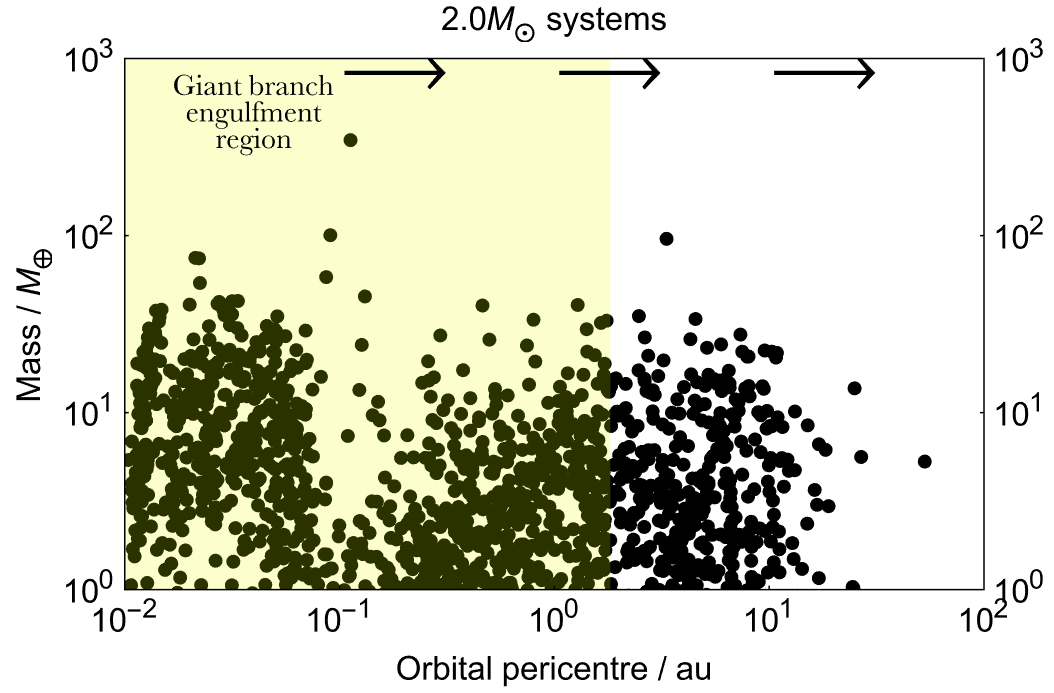}
}
\centerline{}
\centerline{
\includegraphics[width=9.5cm]{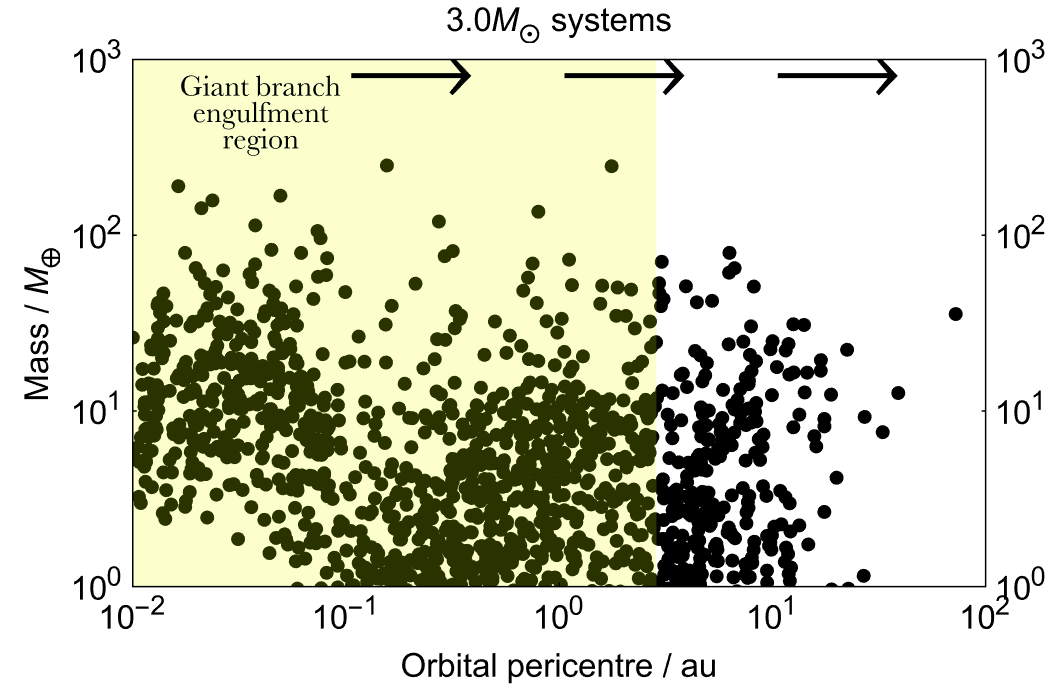}
\includegraphics[width=9.5cm]{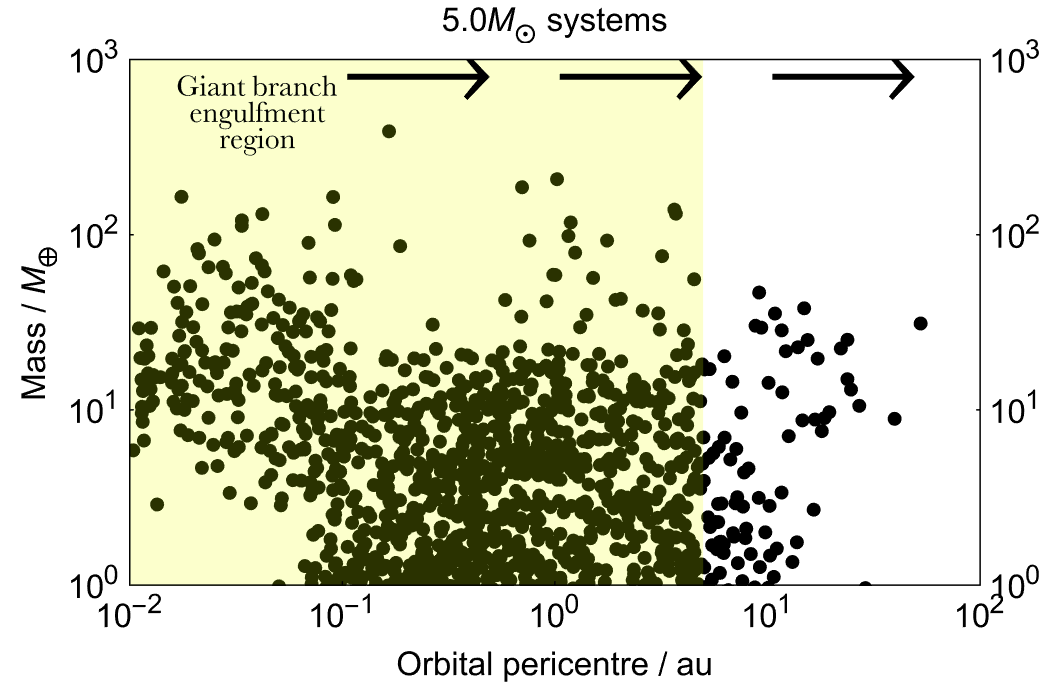}
}
\caption{
{\rev
The planetary-mass objects formed in the simulation ensemble which generated the sub-terrestrial systems that are displayed in Fig. \ref{FigPostMS}. Just as in that figure, here the simulation stopping time in the top two panels is 1~Gyr, in the lower-left panel is 0.4~Gyr, and in the lower-right panel is 0.1~Gyr.
}
}
\label{MassivePostMS}
\end{figure*}

}

\label{lastpage}
\end{document}